\documentclass[aps,prl,twocolumn,nofootinbib,superscriptaddress]{revtex4}
\usepackage{latexsym}
\usepackage{hyperref}
\usepackage{amssymb}
\usepackage{epsfig,amsmath,graphics}
\usepackage{pstricks}
\usepackage{color}
\usepackage{dcolumn}
\usepackage{ulem}
\usepackage{xcolor}
\usepackage{placeins}

\definecolor{linkcolor}{rgb}{0.0, 0.26, 0.67}
\hypersetup{colorlinks=true,
linkcolor=black,
citecolor=linkcolor,
urlcolor=linkcolor,
linktocpage=true
}

\newcommand{\mev}{\text{MeV}}
\newcommand{\eV}{\text{eV}}

\usepackage{comment}
\newcommand{\be}{\begin{equation}}
\newcommand{\ee}{\end{equation}}

\newcommand{\kev}{{\rm keV}}
\newcommand{\Mev}{{\rm MeV}}

\def\bea{\begin{eqnarray}}
\def\eea{\end{eqnarray}}
\def\ltap{\ \raise.3ex\hbox{$<$\kern-.75em\lower1ex\hbox{$\sim$}}\ }
\def\gtap{\ \raise.3ex\hbox{$>$\kern-.75em\lower1ex\hbox{$\sim$}}\ }
\def\lsim{\ \raise.3ex\hbox{$<$\kern-.75em\lower1ex\hbox{$\sim$}}\ }
\def\gsim{\ \raise.3ex\hbox{$>$\kern-.75em\lower1ex\hbox{$\sim$}}\ }

\usepackage{slashed}

\usepackage{color}

\newcommand{\ignore}[1]{}
\newcommand{\beq}{\begin{equation}}
\newcommand{\eeq}{\end{equation}}
\newcommand{\bear}{\begin{eqnarray}}
\newcommand{\eear}{\end{eqnarray}}

\def\mev{\,{\rm MeV}}
\def\kev{\,{\rm keV}}

\def\ev{\,{\rm eV}}

\newcommand{\Veff}{V_{\rm eff}}
\newcommand{\Neff}{N_{\rm eff}}
\newcommand{\dNeff}{\triangle N_{\rm eff}}

\newcommand{\LCDM}{\Lambda{\rm CDM}}
\newcommand{\nuSM}{\nu}

\newcommand{\rhoDS}{\rho_{\rm DS}}
\newcommand{\GammaDS}{\Gamma_{\rm DS}}

\newcommand{\TDS}{T_{\rm DS}}
\newcommand{\TSM}{T_{\nu}}
\newcommand{\nudark}{\ensuremath{\nu_{d}}}
\newcommand{\mdark}{m_{\nu d}}
\newcommand{\ndark}{n_{\rm DS}}
\newcommand{\alphad}{\alpha_{d}}
\newcommand{\Tequil}{T_{\rm equil}}

\begin{document}

\title{Dark Radiation from Neutrino Mixing after Big Bang Nucleosynthesis}

\author{Daniel Aloni}
\affiliation{Physics Department, Boston University, Boston, Massachusetts 02215, USA}
\affiliation{Department of Physics, Harvard University, Cambridge, Massachusetts 02138, USA}
\author{Melissa Joseph}
\affiliation{Physics Department, Boston University, Boston, Massachusetts 02215, USA}
\author{Martin Schmaltz}
\affiliation{Physics Department, Boston University, Boston, Massachusetts 02215, USA}
\affiliation{Center for Cosmology and Particle Physics, Department of Physics, New York University, New York, New York 10003, USA}
\author{Neal Weiner}
\affiliation{Center for Cosmology and Particle Physics, Department of Physics, New York University, New York, New York 10003, USA}
\begin{abstract}

Light dark fermions can mass mix with the Standard Model neutrinos. As a result, through oscillations and scattering, they can equilibrate in the early universe.
Interactions of the dark fermion generically suppress such production at high temperatures but enhance it at later times. 
We find that for a wide range of mixing angles and interaction strengths equilibration with SM neutrinos occurs at temperatures near the dark fermion mass. For masses below an MeV, this naturally occurs after nucleosynthesis and opens the door to a variety of dark sector dynamics with observable imprints on the CMB and large scale structure, and with potential relevance to the tensions in $H_0$ and $S_8$.
\end{abstract}

\maketitle

{\bf Introduction}\quad
The range of redshifts between $z\sim10^9$ and $z\sim 10^3$ correspond to a ``desert" in $\Lambda$CDM. As the temperature cools below the MeV scale where big bang nucleosynthesis (BBN), neutrino decoupling, and $e^+\,e^-$ annihilation take place, no new threshold is reached for almost 6 orders of magnitude until the eV scale where matter-radiation equality, CMB decoupling, and eventually the sum of the neutrino masses can be found.
The $\Lambda$CDM desert originates from the coincidence of a large gap in the mass spectrum of the standard model between the electron mass and the scale of neutrino masses with an unrelated but perfectly overlapping gap between nuclear and atomic binding energies.

Additional dark sectors can have new particles with masses in or below these scales, possibly leading to a rich phenomenology in the desert. 
A minimal extension of the standard model that realizes this has one noninteracting neutral dark fermion \nudark , with mass $\mdark$ in the desert, which mixes with the standard Model (SM) neutrino via a small Dirac mass. A combination of oscillations and weak interaction scattering can easily populate this species for large enough mixing. The relevant rate of this process $\Gamma/H$ peaks near $T\sim 100 \mev \left[\mdark/\kev\right]^{1/3}$~\cite{Dodelson:1993je}, yielding a fully thermalized fermion $\dNeff\approx1$ at BBN for mixing angles  $\sin \theta_0 \gsim 10^{-3}$ and a dark fermion mass $\mdark$ anywhere in the desert. This additional radiation affects BBN and is excluded from the measurements of light element abundances which require $\dNeff|_{T\sim 1\rm MeV}\leq 0.407  \, (95.45\%)$~\cite{Yeh:2022heq}. Smaller mixing angles yield dark fermions which are unthermalized and cosmologically uninteresting as radiation (absent a population from pre-TeV processes), and highly constrained as dark matter. 

However, this minimal picture raises many questions, in particular regarding the origin of the new particle's mass. A natural expectation would be that the mass arises from some dynamics, and there would be other particles and interactions, such as self-interactions, connected to it. The consequences of such an interaction can be significant. Light fermions with large mixings can have their oscillations suppressed in the early universe~\cite{Dasgupta:2013zpn,Hannestad:2013ana,Chu:2015ipa,Cherry:2016jol,Farzan:2019yvo}, changing the cosmological constraints significantly.  In the presence of a self-interaction, regions of parameter space arise where a $\sim \kev$ fermion with small mixings can be dark matter~\cite{Hansen:2017rxr,Johns:2019cwc,DeGouvea:2019wpf,Bringmann:2022aim}. In contrast, absent self-interactions, direct production of such dark matter through weak interactions is excluded by a combination of x-ray data and the presence of small scale structure~\cite{Abazajian:2021zui} (a famous loophole exists when SM neutrinos have chemical potentials and a lepton asymmetry~\cite{Shi:1998km}). Thus it is clear that a dark fermion with interactions is qualitatively different from the ``unnaturally minimal'' scenario of an inert dark state.
Upcoming CMB and LSS observations will probe the $\Lambda$CDM desert, motivating a broader exploration of such models. 

In this Letter we study the equilibration of dark sectors with the SM neutrinos after BBN and neutrino decoupling. Equilibration relies on the dark sector containing at least one neutral fermion which can mix with SM neutrinos and has interactions in the dark sector. 
%
For concreteness, we consider a single dark fermion \nudark\, which mixes with a SM neutrino by an amount $\sin \theta_0$ in vacuum. We assume that $\nudark$ has a self-interaction mediated by a force carrier $\phi$ with $m_\phi \ll \mdark$ and coupling strength $\alphad$. We find two important results: 

\begin{itemize}
	\item The dark sector comes into equilibrium with the neutrinos over a very large parameter space roughly bounded only by $\theta_0^2 \alphad^2 M_{Pl} > \mdark$, allowing mixing angles ranging from 1 to $10^{-13}$. 
	
	\item
	Over most of the parameter space the temperature at which \nudark\, equilibrates is $\alphad$-independent and given by
	\bea
	\Tequil\simeq \mdark \left(\theta_0^2 \frac{M_{Pl}}{\mdark} \right)^{1/5} \ .
	\eea
\end{itemize}

Thus even though the range of allowed values of $\theta_0$ and $\alphad$ is huge, \nudark\, naturally equilibrates at temperatures near $\mdark$, and at most a few orders of magnitude higher, because of the $1/5$ power. 
Consequently, dark sectors with light ($< \mev$) fermions often equilibrate after BBN and are therefore unconstrained by primordial light element abundances. 

\begin{figure}[t]
	\centering
	\includegraphics[width=0.45\textwidth]{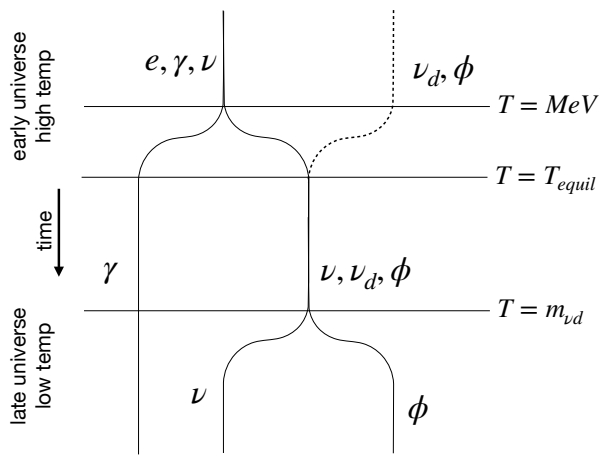}
	\caption{Thermal history of a universe with dark sector thermalization from neutrino mixing after BBN. The dark sector initially has negligible energy density (dashed line). After neutrino decoupling and electron annihilation it equilibrates with the SM neutrinos at $\Tequil$. After $\nudark$ annihilation at $T\sim \mdark$ the SM neutrinos redecouple and free-stream.}
	\label{fig:thermalhistory}
\end{figure}

The simplest thermal history is sketched in Fig. \ref{fig:thermalhistory}. After neutrino decoupling and electron self-annihilation at $T\sim\,$~MeV, the dark sector $\phi$ and $\nu_d$ come into equilibrium with the SM neutrinos. At the lower temperature $T \sim \mdark$, the dark fermions $\nudark$ annihilate away. This causes the SM neutrinos to decouple and become free-streaming again, and the entropy of $\nudark$ is shared between $\phi$ and the SM neutrinos.

Importantly, dark sector equilibration with SM neutrinos {\it after} neutrino decoupling does not change the relativistic energy density because the total energy in neutrinos + dark sector is conserved in the equilibration process. Thus $\Neff$ is unchanged during equilibration, and constraints on $\Neff$ from the CMB and LSS do not {\it a priori} constrain it.

However, if equilibration occurs prior to 100 keV, BBN can be modified. If  $\nu_e$ (rather than $\nu_\mu$ or $\nu_\tau$) equilibrates with $\nudark$, then $\nu_e$ is cooled, suppressing $n\rightarrow p$ conversion. When $T\sim \mdark$ there is a ``step''~\cite{Aloni:2021eaq,Schoneberg:2022grr,Joseph:2022jsf} in the total relativistic energy density (i.e., $\Neff$ increases) as $\nudark$ annihilates away. This can affect BBN as well \cite{Berlin:2019pbq} if it occurs before 100 keV. We leave a detailed study of this for future work.

For later equilibration, BBN is unaffected. However, prior to $T\sim \mdark$, the $\nu-\nudark-\phi$ fluid is tightly coupled. This, combined with the step in $\Neff$ leaves an inevitable imprint on the density perturbations of the universe.

Should other particles have couplings to $\phi$ and $\nudark$, they, too, will come into equilibrium with the SM neutrinos below $T \sim\,$~MeV. As a result, there is a possibility for other interesting dynamics within a dark sector to affect cosmology, such as the thermalization and freeze-out of dark matter, the presence of a second ``step"~\cite{Aloni:2021eaq,Schoneberg:2022grr,Joseph:2022jsf} in the energy density of the dark sector due to the annihilation of additional massive particles into lighter ones. 
Alternatively, in a minimal scenario with $\mdark \lsim$eV, self-interactions in (a portion of) the relativistic energy density may arise only at late times, near recombination. 
Neutrino-dark sector equilibration after BBN thus has very interesting and model-dependent impact on the CMB and structure formation with possible implications for $H_0$ and $S_8$, all of which will be probed by a wide range of upcoming experiments.

\vskip0.5cm
{\bf Interactions and Dark Sector Equilibration} \quad
A generic dark sector which contains a fermion $\nudark$ that mixes with the SM neutrinos can equilibrate with the SM neutrinos very efficiently by the combined effect of  $\nuSM-\nudark$ oscillations and scattering. The relevant formalism is well developed, see~\cite{Barbieri:1989ti,Sigl:1993ctk,Dasgupta:2021ies}.
For simplicity we consider the case of one dark fermion oscillating with one SM neutrino. The rate of conversion of a SM neutrino into a dark fermion can be written as
\begin{align}\label{eq:generic_DW_production}
	\Gamma(E) = \frac{1}{2} \sin^2{2\theta_m} \frac{\Gamma_{\rm int}}{2}~,
\end{align}
where we assume averaging over many oscillations, $\Gamma_{int}$ is the rate of scattering, $\theta_m$ is the in-medium mixing angle between the SM neutrino and the dark fermion and both depend on the incoming neutrino energy $E$. 
The process of dark sector equilibration is the usual competition between the production rate in Eq.~\eqref{eq:generic_DW_production} and Hubble. The mixing angle is generally suppressed by the presence of large diagonal effective thermal masses and thus the overall conversion rate grows rapidly as T declines.

The in-medium mixing angle is given by
\bea 
\label{eq:matter_mixing_angle}
\sin^2{2\theta_m}    = \frac{\sin^2{2\theta_0} }{\left(\cos{2\theta_0} - 2E \triangle \Veff/\triangle m^2 \right) ^2 + \sin^2{2\theta_0} }\ ,
\eea
where $\theta_0$ is the in-vacuum angle that parametrizes the mixing between the SM neutrino and the dark fermion, $\triangle m^2\simeq \mdark^2$ is the mass-squared difference between the two mass eigenstates and is dominated by the dark fermion mass, and $\triangle \Veff=\Veff^{SM}-\Veff^{DS}$. The effective potential of $\nuSM$ from the SM weak interactions is well known~\cite{Dodelson:1993je} and given by $\Veff^{SM} \simeq - c_V G_F^2 T_\nu^4 E$ where $c_V\simeq 22$ (for mixing with $\nu_\mu$ or $\nu_\tau$), and we assume vanishing lepton asymmetry~\cite{Shi:1998km}. The dark sector effective potential arises due to scattering with light particles and a light mediator in the dark thermal bath and can be parametrized as $2E \Veff^{DS}\equiv \alphad T_d^2$~\cite{Chu:2015ipa}.  In what follows we take this as the definition of $\alphad$.
The expression for the effective potential (and dark interaction rate) assumes that the dark sector is self-equilibrated with temperature $T_d$ and vanishing chemical potentials (see discussion below). The exact expression can vary with Dirac/Majorana, internal symmetries and other model dependencies which amount to an overall $O({\rm few})$ rescaling of $\alphad$. The precise mapping onto a specific model Lagrangian is straightforward and not important for our discussion. 
We ignore a possible shift of the scalar expectation value in the thermal background which would change the mass of $\nudark$.

The scattering rate is the sum of the SM weak interaction $\Gamma_{SM}=n_\nu \langle \sigma v \rangle_{SM} =c_\Gamma T_\nu^4 G_F^2 E $ with $c_\Gamma \simeq 0.92$~\cite{Dodelson:1993je}, and the scattering rate of the dark fermions which we parametrize as $\GammaDS =\ndark \langle \sigma v \rangle_{DS} \equiv \kappa\, \alphad^2 T_d^2/E$. This assumes that 
the cross section scales as $\langle \sigma v\rangle_{DS} \simeq \langle \kappa \alphad^2 / E_{\rm CM}^2\rangle_{DS} \simeq \kappa \alphad^2 /(E T_d)$ and $\ndark\propto T_d^3$. Here $\kappa$ is a number greater than one, which allows for the presence of additional dark states which scatter via $\phi$ exchange. For simplicity, we set $\kappa=3$, and in general it would shift the precise region of parameter space but not make it much larger or smaller.

Finally, averaging the conversion rate $\Gamma$ over the thermal distribution of the SM neutrinos approximately replaces $E\rightarrow 3T_\nu$ so that 
\bea
\langle \Gamma\rangle = \frac{ \frac14\sin^2{2\theta_0}(3c_\Gamma T_\nu^5 G_F^2 + \alphad^2 \frac{T_d^2}{T_\nu})}{\left(\cos{2\theta_0} + \alphad \frac{T_d^2}{\mdark^2}+18c_V\frac{G_F^2 T_\nu^6}{\mdark^2} \right)^2+\sin^2{2\theta_0}} \,.
\label{eq:gammarate}
\eea

We can now determine if and when the dark sector equilibrates with the neutrinos by comparing $\Gamma$ with the expansion rate, $H\simeq T_\nu^2/M_{Pl}$.
There are two important limits to consider. First, in the Dodelson-Widrow (DW)~\cite{Dodelson:1993je} limit of vanishing dark sector interactions, $\alphad= 0$, the maximum conversion rate occurs when $G_F T_\nu^3/\mdark\sim 0.1$. This peak temperature is above an MeV so that full equilibration from DW would yield a thermalized dark sector before BBN which is excluded. The dark sector equilibrates if $\Gamma=H$ at the peak; therefore, we obtain the constraint (in the DW limit) that $\theta_0^{2} \mdark M_{Pl}  G_F\lsim100$.

A qualitatively different solution is obtained when the dark sector interactions dominate over the  weak interactions. Then $\langle \Gamma\rangle/H$ grows monotonically with decreasing temperature, and we can solve for the equilibration temperature (when $T_d=T_\nu$) by setting
\bea
\label{eq:Tequileq}
1 \simeq \frac{\langle \Gamma\rangle}{H} 
\simeq \frac{\theta_0^2 \alphad^2 T_\nu}{(1+\alphad \frac{T_\nu^2}{\mdark^2})^2}\frac{M_{Pl}}{T_\nu^2}
\simeq \theta_0^2 \frac{M_{Pl}}{\mdark} \frac{\mdark^5}{T_\nu^5} \ ,
\eea
giving  
$\Tequil=\mdark\, (\theta_0^2  M_{Pl}/\mdark)^{1/5}$.
It is remarkable both that this is independent of $\alphad$ and the dependence on $\theta_0^2 M_{pl}$ is mild because of the $1/5$ power. 
Thus for a very broad range in parameter space the dark sector equilibrates with the neutrinos, and it does so at a temperature which is at most a few orders of magnitude above the dark fermion mass.
This yields the important qualitative result that in the presence of a light ($\ll \Mev$) fermion, the natural equilibration scale is {\it below} the BBN scale, but also above recombination (a similar phenomenology can be achieved in models of neutrinos which couple to a Majoron, and resonantly produce dark matter at late times~\cite{Berlin:2018ztp}.).

\begin{figure}[t]
	\centering
	\includegraphics[width=0.48\textwidth]{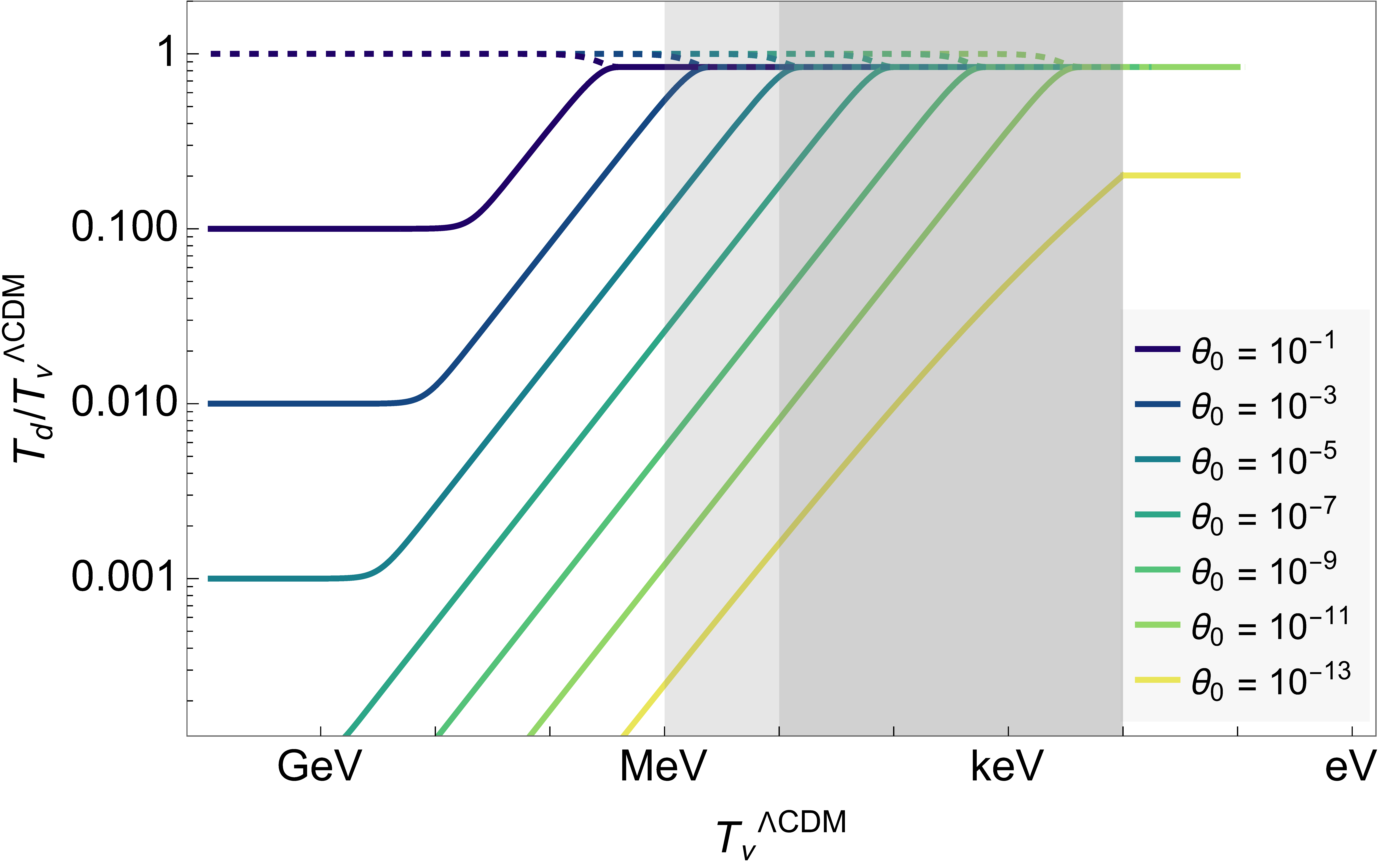}
	\caption{The ratio $T_d/ T_{\nu}^{\LCDM}$
			obtained from solving Eq.~\eqref{eq:Boltzmann} 
			as a function of $T_{\nu}^{\LCDM}$ for an example point with $\alpha_d = 1$, $\mdark=100 \ev$, $ g_*^{\rm DS}/g_*^{\nu} = 1$ and initial dark sector temperature, $T_d$, calculated from Higgs decay. Here $T_{\nu}^{\LCDM}$ the temperature of the active neutrinos in a reference $\Lambda$CDM with no dark sector, where we have neglected changes in $T_{\nu}^{\LCDM}$ from the annihilation of SM particles as they become non-relativistic. The dashed lines correspond to $T_\nuSM/T_\nu^{\LCDM}$ where the small drop shows the approach to equilibrium with the dark sector.
		Equilibration between the sectors occurs when $T_d/ T_{\nu}^{\LCDM} \approx 1$. The dark (light) gray region shows where this occurs after BBN (neutrino decoupling). See text for details.}
	\label{fig:IntregratedRateOverHubble}
\end{figure}

This intuition is borne out by a numerical calculation. Integrating the Boltzmann equations for the phase space distribution functions of dark sector particles against energy and summing over dark sector species we obtain an evolution equation for the total energy density in the dark sector
\bea
\label{eq:Boltzmann}
\frac{d}{d\log a} \left(a^4 \rhoDS\right) = \frac{\langle \Gamma\rangle }{H} a^4
\left(\rho_\nu -\left.\frac{\rho_\nu}{\rhoDS}\right|_{eq.} \rhoDS \right)\ ,
\eea
where $\rhoDS$ is the total energy density in the DS, which we solve numerically. The evolution of the dark sector temperature is shown in Fig.~\ref{fig:IntregratedRateOverHubble}. Details on the calculation of the dark sector temperature evolution are found in Supplemental Materials.

\begin{figure}[t]
	\centering
	\includegraphics[width=0.48\textwidth]{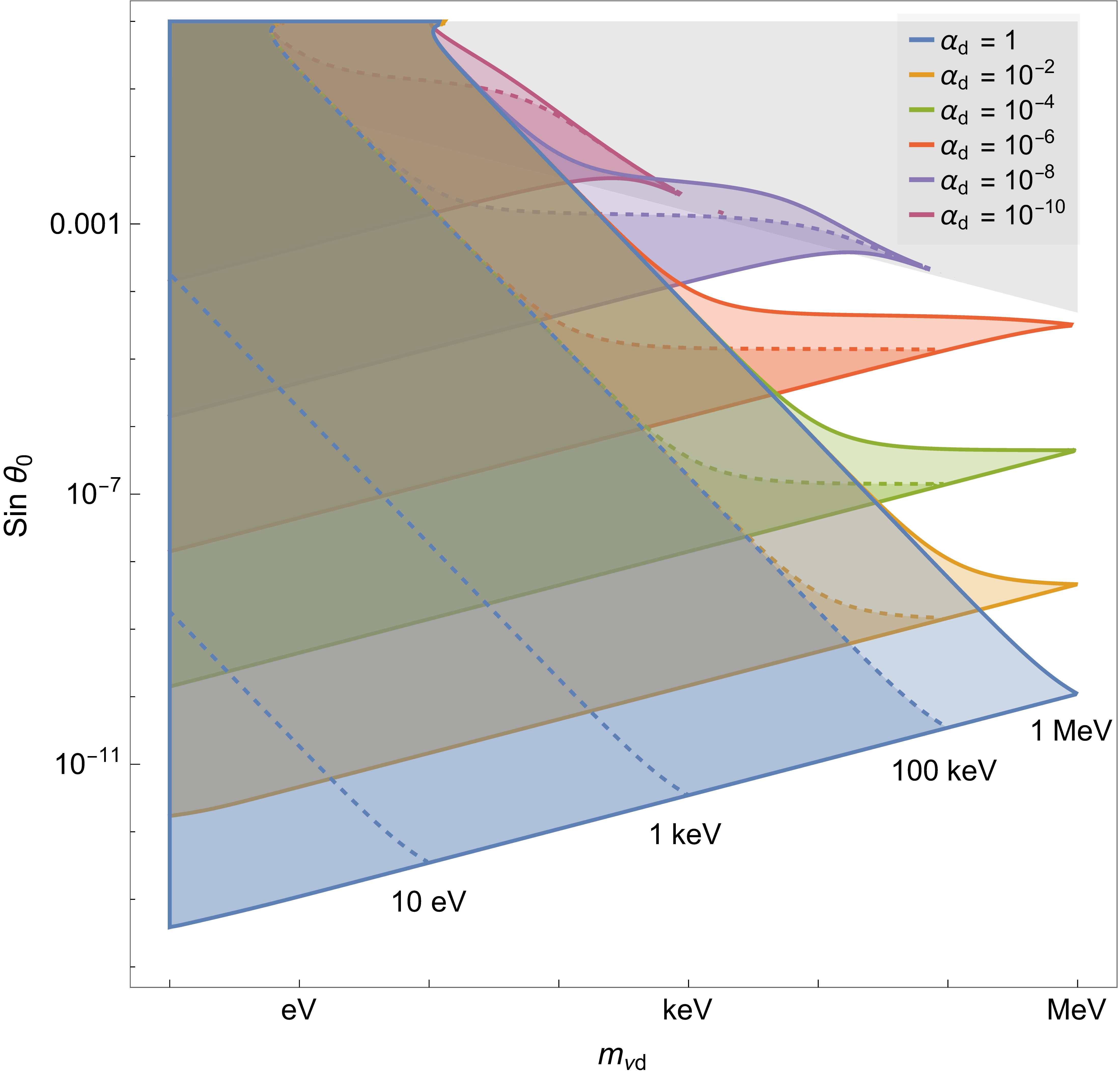}
	\caption{Colored regions indicate the parameter space over which the dark sector comes into equilibrium with the SM neutrinos after BBN, for different values of $\alphad$.  The lower boundary of each region is determined by $\Tequil = \mdark$, while the upper (right) boundary corresponds to equilibration after BBN (dark shaded) or neutrino decoupling (light shaded), i.e. $\Tequil = 100 \kev$ or $= 1 \mev$, respectively. Also shown are contours of fixed equilibration temperatures $\Tequil$ (dashed contours labeled 10 eV, 1 keV) for the $\alpha_d = 1$ case. The gray region shows the parameter space over which equilibration would occur above BBN in absence of dark interactions via Dodelson-Widrow production.}
	\label{fig:ParameterSpace}
\end{figure}

Our primary result is contained in Fig.~\ref{fig:ParameterSpace} which shows the large regions of parameter space where the dark sector comes into equilibrium with the SM neutrinos at some point before $T_\nu=\mdark$ and where equilibration is reached below $T_\nu=\mev$, i.e. after neutrino decoupling and BBN. Note that the small ``fin'' regions on the right of Fig. \ref{fig:ParameterSpace} correspond to parameter space in which $\alphad \Tequil^2/\mdark^2 <1$.
For the purposes of this figure we define the equilibration temperature $\Tequil$ as the temperature at which $\rhoDS$ crosses $\rho_\nu\, g_*^{\rm DS}/g_*^{\nu}$ with $\rhoDS$ obtained from solving Eq.~\eqref{eq:Boltzmann} with the backreaction term omitted.

It is worth noting that because of mixing of the SM neutrinos, for most of parameter space all three SM neutrinos equilibrate with the DS in rapid succession. That only a single SM neutrino equilibrates with the DS can occur for special regions in parameter space. Either the couplings of \nudark\, are tuned such that it only couples to a single SM neutrino mass eigenstate, or the dark parameters are such that equilibration with the first of the SM neutrinos occurs at a temperature just above $\mdark$ so that $\nuSM - \nudark$ conversion shuts off because $\mdark$ is reached before another SM neutrino can equilibrate.

\vskip0.5cm
{\bf Discussion}\quad
One of the simplest extensions of the standard model is to include a massive neutral fermion that mixes with the SM neutrino. It is natural - perhaps expected - that it should come with its own interaction, as well. In the presence of such an interaction, we find that even for very small couplings and mixings, a new eV--- MeV mass fermion is equilibrated with the neutrino bath at a temperature within a few orders of magnitude of its mass, and often much less. Consequently, it typically equilibrates after BBN, leaving no imprint on light element abundances. Its implications for the CMB and LSS, however, can be significant. Once the dark fermion equilibrates at $\Tequil$, a whole series of additional particles can come into equilibrium as well, including dark matter, which can have mass above $\Tequil$, including above an MeV.

Although the equilibration of the dark sector does not immediately increase the energy density in radiation, it can transform some or all of the radiation into an interacting fluid. The associated mass threshold can change the relative amount of relativistic radiation, turn on or off interactions in a dark sector, and provide a basis for equilibrating a broader dark sector which may contain part or all of the dark matter.

\begin{itemize}
	\item{At high values of $100 \ev \lsim \mdark \lsim \mev$, the dark sector equilibrates with neutrinos and then goes through the mass threshold of the dark fermion before the CMB is directly sensitive to the transition. One consequence is the increase in $\Neff$ by $\Delta \Neff = ( (g_*^{UV}/g_*^{IR})^{1/3}-1)N_{eq}$, where $N_{eq}$ is the number of neutrinos that come into equilibrium with the dark sector, and $g_*^{UV} (g_*^{IR})$ is the total number of effective degrees of freedom above (below) the mass threshold, including the thermalizing neutrinos. The relativistic energy below this threshold could be interacting, non-interacting or a combination. }
	
	\item{At intermediate values of $O(1) \ev \lsim \mdark \lsim 100 \ev$, equilibration typically happens before $100\,\eV$, but the mass threshold occurs in a period which is directly probed by the CMB and LSS. This can have important implications for many observables, including $H_0$~\cite{Aloni:2021eaq,Schoneberg:2022grr} and $S_8$~\cite{Joseph:2022jsf}. }
	
	\item{At very low values of $\mdark$, the equilibration can happen below $100\,\eV$, and the signal could appear as a transition of the relativistic energy from free-streaming to strongly interacting. This transition would occur sequentially for the three SM neutrino mass eigenstates and would lead to observable signals in the CMB if it occurred at times near recombination. These implications for the CMB are beyond our scope and warrant their own study.}
\end{itemize}

It is interesting to consider what might be a minimal setup, where a single dark Majorana fermion comes into equilibrium with all three SM neutrinos after BBN, but then annihilates away into a real scalar $\phi$ before the CMB or LSS are directly sensitive. The late universe would have $\Neff \simeq 3.30$ with $ (1-f)\Neff=2.78$ 
free-streaming neutrinos and $ f\Neff=0.53$  interacting particles (arising from $\phi$). Even in this minimal model, the resulting radiation ($\dNeff \simeq 0.26$) is within the bounds from Planck \cite{Planck:2018vyg}, but is well above the sensitivity of Simons Observatory~\cite{SimonsObservatory:2018koc} and CMB-S4~\cite{Abazajian:2019eic}; and the fraction $f=1/(1+3 \cdot 7/4)$ of the ``neutrinos'' that is interacting can be measured from phase shifts of the CMB peaks~\cite{Baumann:2015rya,Pan:2016zla,Kreisch:2019yzn,Blinov:2020hmc, Brinckmann:2020bcn,Brinckmann:2022ajr,RoyChoudhury:2020dmd,RoyChoudhury:2022rva}.

If additional particles couple to $\nudark$ or $\phi$, they, too, will equilibrate at or after $\Tequil$ and the thermal history can be yet richer. If additional light particles are present, then the requirement that $m_\phi \ll \mdark$ is no longer necessary for a viable cosmology. Instead only $m_\phi \ll \Tequil$ is needed for our calculations to hold, and in this case the neutrinos would become free-streaming again at $m_\phi$ rather than $\mdark$. With additional stable particles, dark matter could be produced through thermal processes. For freeze-out, in particular, the dark matter can have masses which are above $\Tequil$, and dark matter would have naturally strong couplings to a radiation bath, at least for some period. In all of these cases, $\Delta \Neff$ can be found simply by an appropriate counting of degrees of freedom in the UV and IR (and intermediate steps, if needed).

In summary, we have considered the thermal history of dark fermions which mix with the SM neutrinos and have self-interactions through a light ($m_\phi \ll \mdark$) mediator. We find that such particles equilibrate at temperatures near their mass, and thus typically at late times. This implies that later universe observables, such as LSS and the CMB are independent probes when compared to BBN for such models. This can have important implications for models attempting to address cosmological tensions. As we look forward to upcoming results from CMB telescopes such as SPT, ACT, Simons Observatory, CMB-S4 as well as studies from LSS measurements KiDS, DES, HSC, and future galaxy surveys with Rubin, Roman, and UNIONS, such models provide an example of natural late-universe phenomena which may have significant impact. Should such particles populate the $\Lambda$CDM desert, these upcoming studies may show striking deviations from  $\Lambda$CDM expectations.

\vskip0.5cm
{\bf Acknowledgments}\quad
We thank Joshua Ruderman and David Dunsky for useful discussions and comments on an early draft. The work of D.A., M.J. and M.S. is supported by the U.S. Department of Energy (DOE) under Award DE-SC0015845. N.W. is supported by NSF under award PHY-2210498, by the BSF under Grants No. 2018140 and 2022287, and by the Simons Foundation. M.S. thanks the CCPP at NYU for their hospitality and support.
\bibliographystyle{utphys}
\bibliography{thermalizationafterbbn}

\providecommand{\href}[2]{#2}\begingroup\raggedright\begin{thebibliography}{10}

\bibitem{Dodelson:1993je}
S.~Dodelson and L.~M. Widrow, ``{Sterile-neutrinos as dark matter},''
  \href{http://dx.doi.org/10.1103/PhysRevLett.72.17}{{\em Phys. Rev. Lett.}
  {\bf 72} (1994)  17--20}, \href{http://arxiv.org/abs/hep-ph/9303287}{{\tt
  arXiv:hep-ph/9303287}}.

\bibitem{Yeh:2022heq}
T.-H. Yeh, J.~Shelton, K.~A. Olive, and B.~D. Fields, ``{Probing physics beyond
  the standard model: limits from BBN and the CMB independently and
  combined},'' \href{http://dx.doi.org/10.1088/1475-7516/2022/10/046}{{\em
  JCAP} {\bf 10} (2022)  046}, \href{http://arxiv.org/abs/2207.13133}{{\tt
  arXiv:2207.13133 [astro-ph.CO]}}.

\bibitem{Dasgupta:2013zpn}
B.~Dasgupta and J.~Kopp, ``{Cosmologically Safe eV-Scale Sterile Neutrinos and
  Improved Dark Matter Structure},''
  \href{http://dx.doi.org/10.1103/PhysRevLett.112.031803}{{\em Phys. Rev.
  Lett.} {\bf 112} (2014) no.~3, 031803},
  \href{http://arxiv.org/abs/1310.6337}{{\tt arXiv:1310.6337 [hep-ph]}}.

\bibitem{Hannestad:2013ana}
S.~Hannestad, R.~S. Hansen, and T.~Tram, ``{How Self-Interactions can Reconcile
  Sterile Neutrinos with Cosmology},''
  \href{http://dx.doi.org/10.1103/PhysRevLett.112.031802}{{\em Phys. Rev.
  Lett.} {\bf 112} (2014) no.~3, 031802},
  \href{http://arxiv.org/abs/1310.5926}{{\tt arXiv:1310.5926 [astro-ph.CO]}}.

\bibitem{Chu:2015ipa}
X.~Chu, B.~Dasgupta, and J.~Kopp, ``{Sterile neutrinos with secret
  interactions\textemdash{}lasting friendship with cosmology},''
  \href{http://dx.doi.org/10.1088/1475-7516/2015/10/011}{{\em JCAP} {\bf 10}
  (2015)  011}, \href{http://arxiv.org/abs/1505.02795}{{\tt arXiv:1505.02795
  [hep-ph]}}.

\bibitem{Cherry:2016jol}
J.~F. Cherry, A.~Friedland, and I.~M. Shoemaker, ``{Short-baseline neutrino
  oscillations, Planck, and IceCube},''
  \href{http://arxiv.org/abs/1605.06506}{{\tt arXiv:1605.06506 [hep-ph]}}.

\bibitem{Farzan:2019yvo}
Y.~Farzan, ``{Ultra-light scalar saving the 3 + 1 neutrino scheme from the
  cosmological bounds},''
  \href{http://dx.doi.org/10.1016/j.physletb.2019.134911}{{\em Phys. Lett. B}
  {\bf 797} (2019)  134911}, \href{http://arxiv.org/abs/1907.04271}{{\tt
  arXiv:1907.04271 [hep-ph]}}.

\bibitem{Hansen:2017rxr}
R.~S.~L. Hansen and S.~Vogl, ``{Thermalizing sterile neutrino dark matter},''
  \href{http://dx.doi.org/10.1103/PhysRevLett.119.251305}{{\em Phys. Rev.
  Lett.} {\bf 119} (2017) no.~25, 251305},
  \href{http://arxiv.org/abs/1706.02707}{{\tt arXiv:1706.02707 [hep-ph]}}.

\bibitem{Johns:2019cwc}
L.~Johns and G.~M. Fuller, ``{Self-interacting sterile neutrino dark matter:
  the heavy-mediator case},''
  \href{http://dx.doi.org/10.1103/PhysRevD.100.023533}{{\em Phys. Rev. D} {\bf
  100} (2019) no.~2, 023533}, \href{http://arxiv.org/abs/1903.08296}{{\tt
  arXiv:1903.08296 [hep-ph]}}.

\bibitem{DeGouvea:2019wpf}
A.~De~Gouv\^ea, M.~Sen, W.~Tangarife, and Y.~Zhang, ``{Dodelson-Widrow
  Mechanism in the Presence of Self-Interacting Neutrinos},''
  \href{http://dx.doi.org/10.1103/PhysRevLett.124.081802}{{\em Phys. Rev.
  Lett.} {\bf 124} (2020) no.~8, 081802},
  \href{http://arxiv.org/abs/1910.04901}{{\tt arXiv:1910.04901 [hep-ph]}}.

\bibitem{Bringmann:2022aim}
T.~Bringmann, P.~F. Depta, M.~Hufnagel, J.~Kersten, J.~T. Ruderman, and
  K.~Schmidt-Hoberg, ``{A new life for sterile neutrino dark matter after the
  pandemic},'' \href{http://arxiv.org/abs/2206.10630}{{\tt arXiv:2206.10630
  [hep-ph]}}.

\bibitem{Abazajian:2021zui}
K.~N. Abazajian, ``{Neutrinos in Astrophysics and Cosmology: Theoretical
  Advanced Study Institute (TASI) 2020 Lectures},''
  \href{http://dx.doi.org/10.22323/1.388.0001}{{\em PoS} {\bf TASI2020} (2021)
  001}, \href{http://arxiv.org/abs/2102.10183}{{\tt arXiv:2102.10183
  [hep-ph]}}.

\bibitem{Shi:1998km}
X.-D. Shi and G.~M. Fuller, ``{A New dark matter candidate: Nonthermal sterile
  neutrinos},'' \href{http://dx.doi.org/10.1103/PhysRevLett.82.2832}{{\em Phys.
  Rev. Lett.} {\bf 82} (1999)  2832--2835},
  \href{http://arxiv.org/abs/astro-ph/9810076}{{\tt arXiv:astro-ph/9810076}}.

\bibitem{Aloni:2021eaq}
D.~Aloni, A.~Berlin, M.~Joseph, M.~Schmaltz, and N.~Weiner, ``{A Step in
  understanding the Hubble tension},''
  \href{http://dx.doi.org/10.1103/PhysRevD.105.123516}{{\em Phys. Rev. D} {\bf
  105} (2022) no.~12, 123516}, \href{http://arxiv.org/abs/2111.00014}{{\tt
  arXiv:2111.00014 [astro-ph.CO]}}.

\bibitem{Schoneberg:2022grr}
N.~Sch\"oneberg and G.~Franco~Abell\'an, ``{A step in the right direction?
  Analyzing the Wess Zumino Dark Radiation solution to the Hubble tension},''
  \href{http://dx.doi.org/10.1088/1475-7516/2022/12/001}{{\em JCAP} {\bf 12}
  (2022)  001}, \href{http://arxiv.org/abs/2206.11276}{{\tt arXiv:2206.11276
  [astro-ph.CO]}}.

\bibitem{Joseph:2022jsf}
M.~Joseph, D.~Aloni, M.~Schmaltz, E.~N. Sivarajan, and N.~Weiner, ``{A Step in
  Understanding the $S_8$ Tension},''
  \href{http://arxiv.org/abs/2207.03500}{{\tt arXiv:2207.03500 [astro-ph.CO]}}.

\bibitem{Berlin:2019pbq}
A.~Berlin, N.~Blinov, and S.~W. Li, ``{Dark Sector Equilibration During
  Nucleosynthesis},'' \href{http://dx.doi.org/10.1103/PhysRevD.100.015038}{{\em
  Phys. Rev. D} {\bf 100} (2019) no.~1, 015038},
  \href{http://arxiv.org/abs/1904.04256}{{\tt arXiv:1904.04256 [hep-ph]}}.

\bibitem{Barbieri:1989ti}
R.~Barbieri and A.~Dolgov, ``{Bounds on Sterile-neutrinos from
  Nucleosynthesis},''
  \href{http://dx.doi.org/10.1016/0370-2693(90)91203-N}{{\em Phys. Lett. B}
  {\bf 237} (1990)  440--445}.

\bibitem{Sigl:1993ctk}
G.~Sigl and G.~Raffelt, ``{General kinetic description of relativistic mixed
  neutrinos},'' \href{http://dx.doi.org/10.1016/0550-3213(93)90175-O}{{\em
  Nucl. Phys. B} {\bf 406} (1993)  423--451}.

\bibitem{Dasgupta:2021ies}
B.~Dasgupta and J.~Kopp, ``{Sterile Neutrinos},''
  \href{http://dx.doi.org/10.1016/j.physrep.2021.06.002}{{\em Phys. Rept.} {\bf
  928} (2021)  1--63}, \href{http://arxiv.org/abs/2106.05913}{{\tt
  arXiv:2106.05913 [hep-ph]}}.

\bibitem{Berlin:2018ztp}
A.~Berlin and N.~Blinov, ``{Thermal neutrino portal to sub-MeV dark matter},''
  \href{http://dx.doi.org/10.1103/PhysRevD.99.095030}{{\em Phys. Rev. D} {\bf
  99} (2019) no.~9, 095030}, \href{http://arxiv.org/abs/1807.04282}{{\tt
  arXiv:1807.04282 [hep-ph]}}.

\bibitem{Planck:2018vyg}
{\bf Planck} Collaboration, N.~Aghanim {\em et al.}, ``{Planck 2018 results.
  VI. Cosmological parameters},''
  \href{http://dx.doi.org/10.1051/0004-6361/201833910}{{\em Astron. Astrophys.}
  {\bf 641} (2020)  A6}, \href{http://arxiv.org/abs/1807.06209}{{\tt
  arXiv:1807.06209 [astro-ph.CO]}}. [Erratum: Astron.Astrophys. 652, C4
  (2021)].

\bibitem{SimonsObservatory:2018koc}
{\bf Simons Observatory} Collaboration, P.~Ade {\em et al.}, ``{The Simons
  Observatory: Science goals and forecasts},''
  \href{http://dx.doi.org/10.1088/1475-7516/2019/02/056}{{\em JCAP} {\bf 02}
  (2019)  056}, \href{http://arxiv.org/abs/1808.07445}{{\tt arXiv:1808.07445
  [astro-ph.CO]}}.

\bibitem{Abazajian:2019eic}
K.~Abazajian {\em et al.}, ``{CMB-S4 Science Case, Reference Design, and
  Project Plan},'' \href{http://arxiv.org/abs/1907.04473}{{\tt arXiv:1907.04473
  [astro-ph.IM]}}.

\bibitem{Baumann:2015rya}
D.~Baumann, D.~Green, J.~Meyers, and B.~Wallisch, ``{Phases of New Physics in
  the CMB},'' \href{http://dx.doi.org/10.1088/1475-7516/2016/01/007}{{\em JCAP}
  {\bf 01} (2016)  007}, \href{http://arxiv.org/abs/1508.06342}{{\tt
  arXiv:1508.06342 [astro-ph.CO]}}.

\bibitem{Pan:2016zla}
Z.~Pan, L.~Knox, B.~Mulroe, and A.~Narimani, ``{Cosmic Microwave Background
  Acoustic Peak Locations},''
  \href{http://dx.doi.org/10.1093/mnras/stw833}{{\em Mon. Not. Roy. Astron.
  Soc.} {\bf 459} (2016) no.~3, 2513--2524},
  \href{http://arxiv.org/abs/1603.03091}{{\tt arXiv:1603.03091 [astro-ph.CO]}}.

\bibitem{Kreisch:2019yzn}
C.~D. Kreisch, F.-Y. Cyr-Racine, and O.~Dor\'e, ``{Neutrino puzzle: Anomalies,
  interactions, and cosmological tensions},''
  \href{http://dx.doi.org/10.1103/PhysRevD.101.123505}{{\em Phys. Rev. D} {\bf
  101} (2020) no.~12, 123505}, \href{http://arxiv.org/abs/1902.00534}{{\tt
  arXiv:1902.00534 [astro-ph.CO]}}.

\bibitem{Blinov:2020hmc}
N.~Blinov and G.~Marques-Tavares, ``{Interacting radiation after Planck and its
  implications for the Hubble Tension},''
  \href{http://dx.doi.org/10.1088/1475-7516/2020/09/029}{{\em JCAP} {\bf 09}
  (2020)  029}, \href{http://arxiv.org/abs/2003.08387}{{\tt arXiv:2003.08387
  [astro-ph.CO]}}.

\bibitem{Brinckmann:2020bcn}
T.~Brinckmann, J.~H. Chang, and M.~LoVerde, ``{Self-interacting neutrinos, the
  Hubble parameter tension, and the cosmic microwave background},''
  \href{http://dx.doi.org/10.1103/PhysRevD.104.063523}{{\em Phys. Rev. D} {\bf
  104} (2021) no.~6, 063523}, \href{http://arxiv.org/abs/2012.11830}{{\tt
  arXiv:2012.11830 [astro-ph.CO]}}.

\bibitem{Brinckmann:2022ajr}
T.~Brinckmann, J.~H. Chang, P.~Du, and M.~LoVerde, ``{Confronting interacting
  dark radiation scenarios with cosmological data},''
  \href{http://arxiv.org/abs/2212.13264}{{\tt arXiv:2212.13264 [astro-ph.CO]}}.

\bibitem{RoyChoudhury:2020dmd}
S.~Roy~Choudhury, S.~Hannestad, and T.~Tram, ``{Updated constraints on massive
  neutrino self-interactions from cosmology in light of the $H_0$ tension},''
  \href{http://dx.doi.org/10.1088/1475-7516/2021/03/084}{{\em JCAP} {\bf 03}
  (2021)  084}, \href{http://arxiv.org/abs/2012.07519}{{\tt arXiv:2012.07519
  [astro-ph.CO]}}.

\bibitem{RoyChoudhury:2022rva}
S.~Roy~Choudhury, S.~Hannestad, and T.~Tram, ``{Massive neutrino
  self-interactions and inflation},''
  \href{http://dx.doi.org/10.1088/1475-7516/2022/10/018}{{\em JCAP} {\bf 10}
  (2022)  018}, \href{http://arxiv.org/abs/2207.07142}{{\tt arXiv:2207.07142
  [astro-ph.CO]}}.

\end{thebibliography}\endgroup

\clearpage
\newpage

\onecolumngrid

\begin{center}

	{ \it \large Supplemental Material}\\

\end{center}
\onecolumngrid
\setcounter{equation}{0}
\setcounter{figure}{0}
\setcounter{table}{0}
\setcounter{section}{0}
\setcounter{page}{1}
\makeatletter
\renewcommand{\theequation}{S\arabic{equation}}
\renewcommand{\thefigure}{S\arabic{figure}}
\renewcommand{\thetable}{S\arabic{table}}
\newcommand\ptwiddle[1]{\mathord{\mathop{#1}\limits^{\scriptscriptstyle(\sim)}}}
\section{Details on the Equilibration of the Dark Sector }

Integrating the Boltzmann equations for the phase space distribution functions of dark sector particles against energy and summing over dark sector species we obtain an evolution equation for the total energy density in the dark sector
\bea
\label{eq:Boltzmann1}
\frac{d}{d\log a} \left(a^4 \rhoDS\right) = \frac{\langle \Gamma\rangle }{H} a^4
\left(\rho_\nu -\left.\frac{\rho_\nu}{\rhoDS}\right|_{eq.} \rhoDS \right)\ ,
\eea

where $\rhoDS$ is the total energy density in the DS including $\nudark, \phi$ and any other (relativistic) particles that the DS may have. We assume that the dark sector starts out cold and initial dark abundances arise from Higgs decay and weak interactions which are determined from the values $\theta_0$ and $\mdark$. A small additional population below a thermal abundance does not significantly change the results. On the right-hand side of Eq.~\eqref{eq:Boltzmann1} is a source term corresponding to the influx of energy into the dark sector from neutrino conversions and a back reaction term accounting for outflux of energy from $\nudark \rightarrow \nuSM$ conversions. The back reaction term is negligible until the dark sector and the SM neutrinos are close to equilibrium and ensures the correct approach to equilibrium when the two terms cancel. The evolution of the SM neutrino temperature can also be found by solving the coupled equation for $\rho_\nu$
	\bea
	\label{eq:Boltzmann2}
	\frac{d}{d\log a} \left(a^4 \rho_\nu\right) = - \frac{\langle \Gamma\rangle }{H} a^4
	\left(\rho_\nu -\left.\frac{\rho_\nu}{\rhoDS}\right|_{eq.} \rhoDS \right)\ .
	\eea
Note that interactions of $\nudark$ with all other dark sector particles redistribute energy within the dark sector but do not contribute to the evolution of the total energy density of the DS in (\ref{eq:Boltzmann1}). 

The assumption that the DS is self-thermalized which allowed us to write $\langle \Gamma \rangle$ as a function of the dark sector temperature is not necessarily true. However, the DS always reaches kinetic equilibrium before it equilibrates with the neutrinos and in most of parameter space number-changing interactions in the DS also erase any chemical potentials before equilibration with the neutrinos. For simplicity, we assume that the DS self-thermalizes rapidly in Figs.~\ref{fig:IntregratedRateOverHubble} and \ref{fig:ParameterSpace}. For small $\alphad$ this may require additional interactions which could be in $V(\phi)$ or involve additional dark sector particles. 

For Fig.~\ref{fig:IntregratedRateOverHubble}, we solve Eq.~\eqref{eq:Boltzmann1} and Eq.~\eqref{eq:Boltzmann2} numerically for $\rhoDS$ and $\rho_\nu$ and obtain $\TDS, \TSM$ using the relations $\rhoDS = g_*^{\rm DS} \TDS^4$ and $\rhoDS = g_*^{\rm \nu} \TSM^4$ neglecting any changes in $H$ from the annihilation of SM particles as they become non-relativistic. As seen in Eq.~\eqref{eq:Tequileq}, for the part of parameter space considered in this Letter, the equilbration temperature is relatively insensitive to processes at $T \gg $ MeV . The evolution of the DS energy density depends on the model parameters. Fig.~\ref{fig:IntregratedRateOverHubble} shows the DS temperature evolution for different values of $\theta_0$ with fixed $\mdark=100$ eV and $\alphad=1$. The Figure shows $T_d/T_\nu^{\LCDM}$ as a function of a reference $\LCDM$ neutrino temperature in solid lines. As the dark sector comes into equilibrium with the SM neutrinos, $T_d$ merges with the SM neutrino temperature, $T_\nuSM$. The effect of the back reaction term in Eq.~\eqref{eq:Boltzmann1} on the SM neutrino temperature is shown in Fig.~\ref{fig:IntregratedRateOverHubble} with dashed lines that correspond to $T_\nuSM/T_\nu^{\LCDM}$. From Fig.~\ref{fig:IntregratedRateOverHubble} it is visible that the equilibration of the dark sector is IR dominated. This justifies our simplification of neglecting changes in $g_{*S}$ in the calculation of $H$. Moreover, from Fig.~\ref{fig:IntregratedRateOverHubble} it can also be seen that the effect of the backreaction is subleading and only relevant very close to equilibration therefore we omit the backreaction effect and solve only Eq.~\eqref{eq:Boltzmann1}, assuming $\TSM \sim 1/a$, when producing Fig.~\ref{fig:ParameterSpace}. Equilibration in the gray region corresponds to equilibration between BBN ($T_\nu \simeq \mev$) and $T_\nu=\mdark$. If equilibration is not reached before $T_\nu=\mdark$ it is never reached because the interaction $\langle \Gamma\rangle/H $ rapidly shuts off for $T_\nu < \mdark$.
We note that $\alphad \simeq 1$ may require higher orders in perturbation theory for precise predictions. Nevertheless, we use it as an example because it allows the largest range of angles $\theta_0$ to equilibrate, see Fig.~\ref{fig:ParameterSpace}.

\end{document}